# Evidence of trion-libron coupling in chirally adsorbed single molecules

Jiří Doležal[1,2]*, Sofia Canola[1], Prokop Hapala[1], Rodrigo Cezar de Campos Ferreira[1], Pablo Merino[3,4], Martin Švec[1,5]*

[1] Institute of Physics, Czech Academy of Sciences; Cukrovarnická 10/112, CZ16200 Praha 6, Czech Republic

[2] Faculty of Mathematics and Physics, Charles University; Ke Karlovu 3, CZ12116 Praha 2, Czech Republic

[3] Catalan Institute of Nanoscience and Nanotechnology (ICN2), CSIC and BIST, Campus UAB, Bellaterra, E08193 Barcelona, Spain

[4] Instituto de Ciencia de Materiales de Madrid; CSIC, Sor Juana Inés de la Cruz 3, E28049 Madrid, Spain

[5] Institute of Organic Chemistry and Biochemistry, Czech Academy of Sciences; Flemingovo náměstí 542/2. CZ16000 Praha 6, Czech Republic

## Abstract

Interplay between motion of nuclei and excited electrons in molecules plays a key role both in biological and artificial nanomachines. Here we provide a detailed analysis of coupling between quantized librational modes (librons) and charged excited states (trions) on single phthalocyanine dyes adsorbed on a surface. By means of tunnelling electron-induced electroluminescence, we identify libronic progressions on a µeV energy range in spectra of chirally adsorbed phthalocyanines, which are otherwise absent from spectra of symmetrically adsorbed species. Experimentally measured libronic spectra match very well the theoretically calculated libron eigenenergies and peak intensities (Franck-Condon factors) and reveal an unexpected depopulation channel for the zero libron of the excited state that can be effectively controlled by tuning the size of the nanocavity. Our results showcase the possibility of characterizing the dynamics of molecules by their low-energy molecular modes using µeV-resolved tip-enhanced spectroscopy.

Coupling between excited electronic states and nuclear motion is an essential mechanism for conversion between optical, mechanical and chemical forms of energy in nanosystems. It plays a crucial role in biological processes such as photosynthesis[1] or light-sensitive proteins in eyes[2,3], as well as in artificial molecular motors[4,5] or organic solar cells[6,7]. Coupling of optical states with vibrations was also found to have major importance for understanding the quantum coherence in coupled fluorophore systems in biological context[1]. On one hand, vibrations provide an important decoherence mechanism, but on the other hand their time evolution may mimic quantum beating[8], which can be particularly relevant for eventual development of optical quantum computers[9].

Frustrated rotations (librations) represent a particular type of periodic molecular motion in which the molecule performs a torsional oscillation when subjected to external stimuli and constraints that restrict its orientation. Despite their efficient coupling to electronic transitions, librations have largely eluded spectroscopic detection because of naturally relatively small energy differences between their quantized levels, making it difficult to derive any characteristic parameters from the spectra, especially in large ensembles of molecules. Moreover, detailed study of these phenomena directly (e.g. in biological systems) is complicated because they become easily obscured by stochastic thermal motion, the effect of solvents and by a generally limited control over the nanoscopic environment of the chromophores. Therefore, performing experiments in well-controlled environments on the single-molecule level is a key to advancing our fundamental understanding of molecular librations as well as for the development of advanced nanomachines and nanodevices.

Recent advances in tip-enhanced single-molecule spectroscopy permit to overcome the limitations of traditional ensemble-based spectroscopies and study neutral and charged excited states and their coupling to vibrations on single molecules. Scanning tunnelling microscope-induced electroluminescence[10,11,12,13,14,15,16,17,18] (STM-EL), photoluminescence[19,20] (STM-PL) and tip-enhanced Raman scattering[21,22,23] (TERS) methodologies operating at cryogenic temperatures can localize and amplify the interaction of electromagnetic radiation with a molecule located in the plasmonic nanocavity formed between the scanning probe and a metal sample by many orders of magnitude. Using these approaches one can study individual photoactive molecules without the influence of stochastic thermal fluctuations or the presence of solvents, and map the electron transitions in the optical near-field with submolecular resolution. This resolution is orders of magnitude higher than what can be achieved with in-solution spectroscopy and opens a new window to determine the mechanisms governing the photophysics of molecular systems.

Here we apply STM-EL spectroscopy to investigate coupling between charged excited electronic states (trions) and quantized librations (librons) of zinc, magnesium and free-base phthalocyanine molecules (ZnPc, MgPc and $H_2Pc$ respectively), adsorbed on sodium chloride (NaCl). Phthalocyanines are structurally similar to biological fluorophores (e.g. chlorophyll), therefore their interaction with the crystalline substrate provides a convenient controllable model for more complex interactions *in vivo*. Also, frustrated rotations of phthalocyanines on surfaces have been proposed as a model for molecular rotors and switches[24,25] for their ability to jump between various adsorption geometries on the surface upon electronic or mechanical excitation. We exploit the fact that adsorbed fluorophores exposed to electric fields in the nanocavity show propensity to charge and emit from excited trion states that generally manifest substantially narrower lineshapes, compared to the emission peaks of the neutral excited states[26,27]. This enables high spectroscopic resolution suitable for studying the fine structure arising from the trion-libron coupling, which we rationalize using time-dependent density functional theory (TD-DFT) calculations and the Franck-Condon principle. Using this approach we can precisely extract parameters of the potential energy landscape of the systems in their respective ground and excited states, the libration eigenenergies and estimate the probability distribution of the librons in the excited states. We can establish a general correlation between adsorption configuration (chiral vs. non-chiral) and the spectral profile, determined from the intensity of Franck-Condon factors of the transitions.

## High-resolution STM-EL spectra of single phthalocyanine adsorbates

Zinc-phthalocyanine (ZnPc) and Magnesium-phthalocyanine (MgPc) adsorb centred on the $Cl^-$ site of 2-5 monolayers (ML) of NaCl on Ag(111) and manifest a characteristic 16-lobe appearance in the occupied-state STM images measured at -2.8 V (insets in Fig. 1b-c). This appearance is the result of averaging between two geometrically equivalent metastable chiral adsorption configurations that rapidly switch upon injection of electrons[12,13,16,28,29]. The motion between these two configurations is represented by the larger grey arrow in Fig. 1a. In contrast, free-base phthalocyanine ($H_2Pc$) adsorbs centred above the $Na^+$ site and exhibits an 8-lobe pattern in the STM image (inset in Fig. 1d). Here, the apparent symmetry is due to averaging over the $H_2Pc$ tautomers with different configurations of the central two H atoms[30]. STM-EL spectra (see Figure 1b-c) acquired at bias voltages below -2.6 V on the molecular lobes of ZnPc and MgPc show distinct emission peaks, corresponding to neutral exciton (Q) and of the cation trion ($Q^+$). For $H_2Pc$ in the neutral state, the excited states with electric transition dipoles oriented along the *x* and *y* axes of the molecule ($Q_x$ and $Q_y$) are not equivalent, which results in degeneracy lifting and two observable excitonic lines: $Q_x$ and $Q_y$, the former having lower energy and higher intensity. Interestingly, for $H_2Pc$ in the cationic state only a single trion peak is detected. Based on quantum-chemical calculations (Extended Data Fig. 1 and Extended Data Table 1), we assign it to the $Q_y^+$ exciton, as the $Q_x^+$ is predicted to be of about 200 meV higher in energy[18]. In all three species, the vibronic progressions associated with the emission are observed in the energy range between their respective neutral exciton and trion peaks, most prominently in the case of $H_2Pc$ (Fig.1d), where the intensities of vibrational contributions are amplified by the effect of nanocavity.

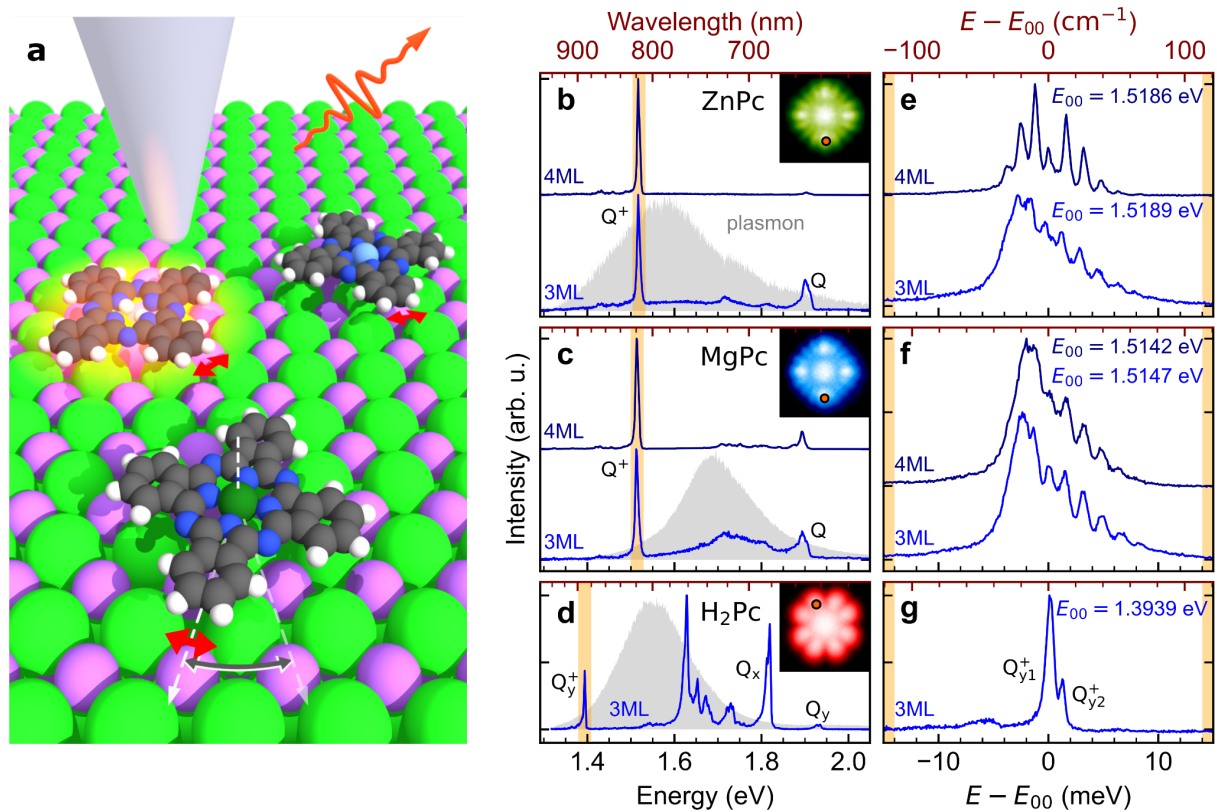

*Fig.1:* a) Scheme of the STM-EL measurement of the $MgPc$, $H_2Pc$, $ZnPc$ molecules (from top to bottom, respectively), performing in-plane librations on the surface of NaCl/Ag. The red arrow denotes small-energy librations while the grey arrow represents switching between two degenerate chiral equilibrium adsorption positions. b-d) Overview STM-EL spectra of the ZnPc, MgPc and $H_2Pc$ at -2.8 V, 100 pA, showing the neutral (Q, $Q_x$, $Q_y$) and cation ($Q^+$, $Q_y^+$) emission fingerprints. Grey-filled spectra on the background of each panel are the responses of the nanocavities measured on a clean Ag(111) surface at 2.5 V, 1 nA. e-g) Spectra measured at the same bias and tunnelling current with 400 µeV resolution on the ZnPc, MgPc and $H_2Pc$ cations, respectively, evidencing the fine structure present in the first two cases. The scale is given relative to the central peaks in the spectral manifold, which manifest lower intensity with respect to their neighbours. The reference energies $E_{00}$ are set to the central peaks with low intensity in each spectrum.

The full width at half maximum (FWHM) of the ZnPc and MgPc neutral Q peaks are typically 8-20 meV (Extended Data Fig. 2, ref [31]) depending on the exact NaCl thickness, tip-sample separation[32] and nanocavity plasmon-exciton matching[32,33]. Conversely, the $Q_x$, $Q_y$ linewidths of neutral $H_2Pc$ are narrower - as low as 4 meV[20,33], but still several orders of magnitude larger than a homogeneous broadening on a comparable system at 6 K[34,35]. The width of the $ZnPc^+$ and $MgPc^+$ trion envelopes are typically in the range 5-7 meV, however, high-resolution spectra reveal a rich fine structure. When measured at 400 µeV resolution, the spectra manifest a manifold of narrow, nearly regularly spaced peaks (Figs. 1e,f), where typically the peak in the centre of the manifold shows less intensity than its neighbours. Surprisingly, such fine structure is absent in the high-resolution spectrum of the $H_2Pc^+$ trion (in Fig. 1g) which shows only a main line, accompanied by a second minor component, originating from the second tautomer.

**Theoretical model of the librations and fitting of the spectra**

The spectral fingerprints of the $ZnPc^+$ and $MgPc^+$ trions comprising multiple peaks indicate an efficient coupling between the molecular librations and electronic transitions. In order to test this notion and to elucidate the observed complex spectral features, we first perform TD-DFT calculations of the total energy $E$ dependence on azimuthal angle $\phi$ for the molecule cations in their doublet ground and first excited states (denoted as $D_0$, $D_1$). The results of the $E(\phi)$ calculations are summarized in Fig. 2. We find that $ZnPc^+$ and $MgPc^+$ in the ground and trion states have double-well energy landscapes (consistent with refs [28,36]) with equilibrium azimuthal angle $\phi_0 \approx \pm 15°$ (Figs.2a-b), separated by a barrier of ~200 meV. The $H_2Pc^+$ energy landscape consists of a single well with the lowest energy configuration at $\phi_0 = 45°$ (Fig.2c). For the transitions $D_0 \rightarrow D_1$ in $ZnPc^+$ and $MgPc^+$, parabolic fits of $E(\phi)$ around the equilibrium angle (Fig. 2d and Extended Data Table 2) quantify a shift in the equilibrium angle $\Delta\phi_0$ of about 0.3° and a change in the stiffnesses of a few percent ($k_0$ and $k_1$ respectively). The nonzero $\Delta\phi_0$ is a result of the different asymmetry of substrate electrostatic interaction acting between the NaCl substrate and the ground and excited states of the chirally adsorbed molecules. $H_2Pc^+$ also shows a comparable change in the stiffness of the potential well upon excitation, but at the same time it does not rotate its equilibrium configuration (*i.e.* $\Delta\phi_0 = 0$), due to its symmetrical adsorption geometry with respect to NaCl.

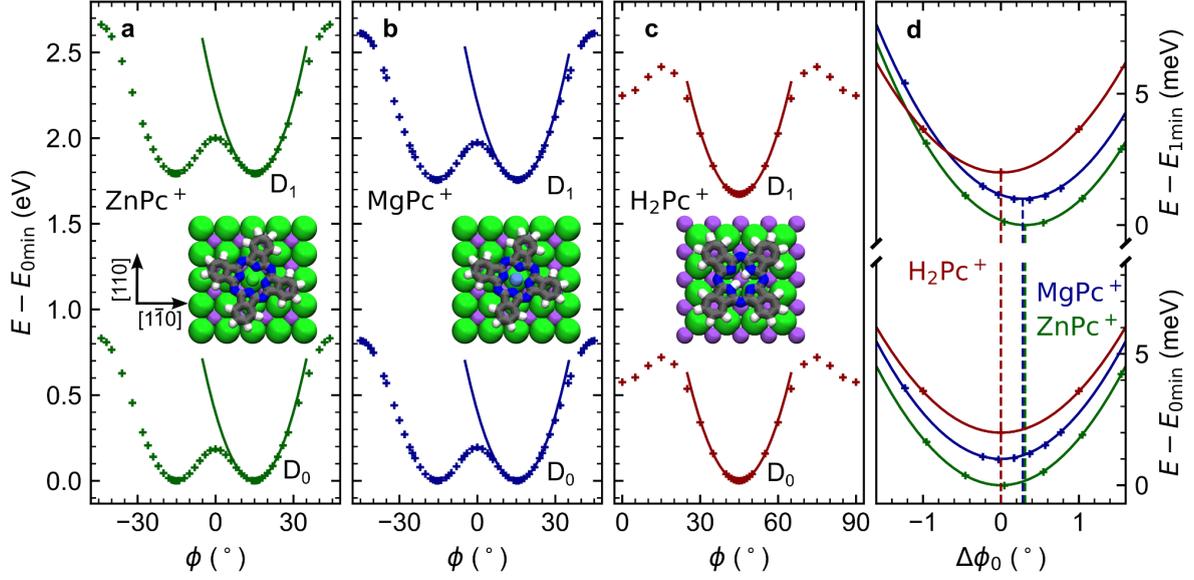

**Fig.2:** *Total energy (as a difference from the energy of the minima $E_{0min}$) as a function of rotation by angle $\phi$ for the ground and excited states of a) $ZnPc^+$, b) $MgPc^+$ and c) $H_2Pc^+$. The computed energy is plotted with crosses and the corresponding parabolic fits around the local minima with solid lines. The insets show the schematic models of the respective ground state cations in their equilibrium positions. The angle $\phi$ is defined as between the molecule x-axes (crossing two opposing isoindole groups along N - N atom direction) and the [110] of NaCl. d) Detailed comparison of the potential well minima of the three cationic chromophores (computed as a difference with the energy of their respective minima $E_{0min}$ or $E_{1min}$) as a function of the shift in the equilibrium angle positions $\Delta\phi_0$ between the ground and excited states. $MgPc^+$ and $H_2Pc^+$ ground and excited state are vertically offset by increments of 1 meV for clarity. Note the zero $\Delta\phi_0$ for $H_2Pc^+$, dictated by the symmetry of the system.*

The calculations show that for energies well below the barriers, the librating molecules can be treated as harmonic torsional oscillators. Using a harmonic molecule-surface interaction potential $V(\phi)$, defined by the stiffness according to the electronic state ($k_0$ or $k_1$) and moment of inertia $J$, we can determine the librational eigenenergies ($\varepsilon_i$) and corresponding wavefunctions $\psi_i$ in the ground and excited state as a function of $\phi$, by solving numerically the 1D Schrodinger equation

$$[V(\phi) - \frac{\hbar^2}{2J}\frac{\partial^2}{\partial\phi^2}]\psi_i(\phi) = \varepsilon_i\psi_i(\phi) \qquad (1)$$

For the description of trion-libron coupling we employ the Franck-Condon principle, in analogy with the description of vertical vibronic transitions between excited and ground states in molecular systems (see Fig. 3a,b). Accordingly, the intensity $I_{mn}$ of a libronic peak associated with the decay from the electronically excited librational state *m to* the electronic ground librational state *n* can be approximated as:

$$I_{mn} \propto \mu^2 w_m <\psi_m|\psi_n>^2 \qquad (2)$$

where μ is the electronic transition dipole moment between the ground and excited electronic states, and $<\psi_m|\psi_n>^2$ is the squared modulus of the overlap integral between the wavefunctions ψ of librational states, *i.e.* the Franck-Condon factors. $w_m$ describes the probability of the system to be in the initial state *m* of energy $\varepsilon_m$ at the moment of emission, and we model it by an exponential distribution, using an effective temperature $T_{eff}$ of the system:

$$w_m = \frac{e^{-\varepsilon_m/k_B T_{eff}}}{Z} \qquad (3)$$

where $k_B$ is the Boltzmann constant and $Z = \sum_i e^{\varepsilon_i/k_B T_{eff}}$ is the partition function, whose sum runs over all libronic states *i*.

Next, we simulate the emission spectrum as the sum of all emission lines (up to *m,n = 50*) with energy shifts $\varepsilon_m - \varepsilon_n$ and intensity $I_{mn}$, convolved with a dressing Gaussian (Lorentzian for $H_2Pc^+$, see Methods) function to account for the additional spectral broadening, with a parametrically imposed full width at half maximum (γ).

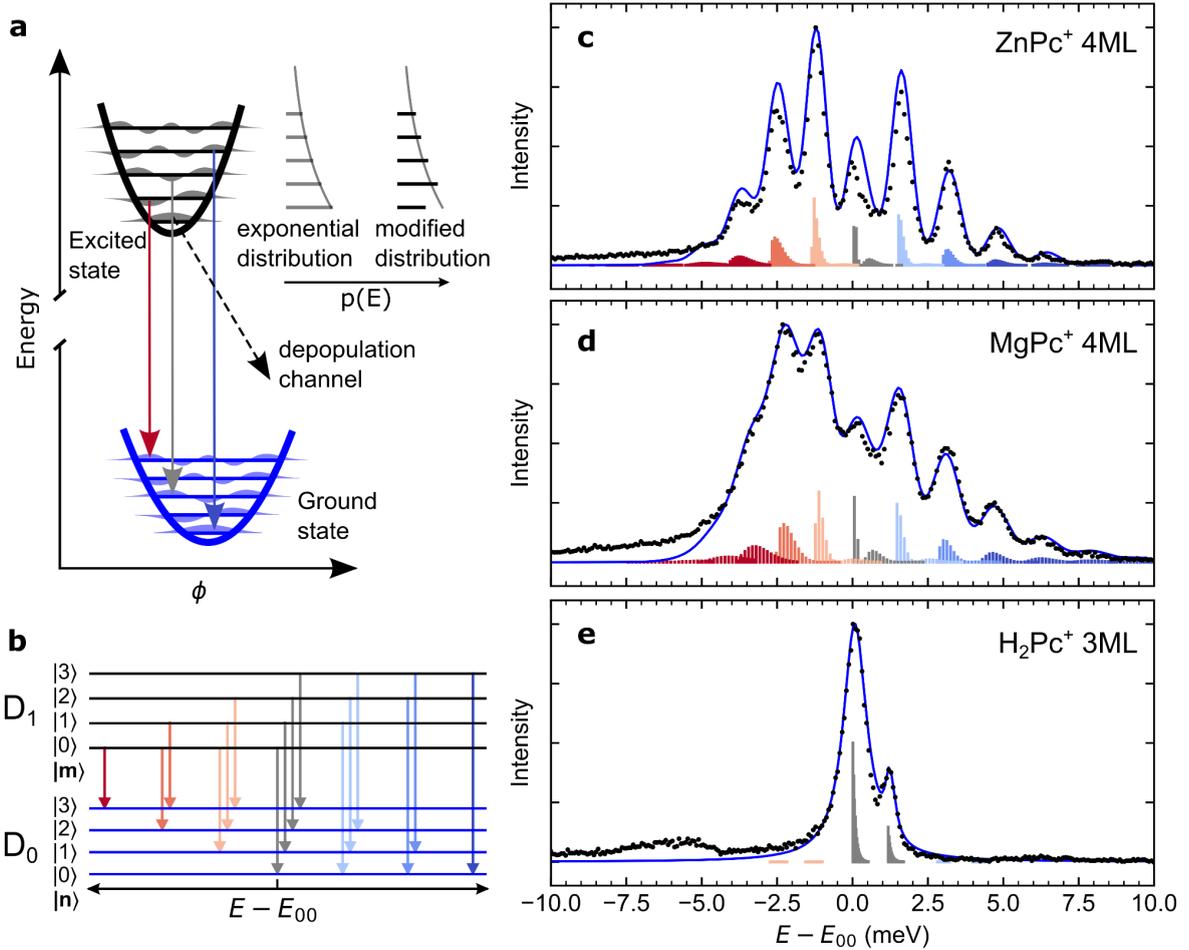

***Fig.3:*** *a) Scheme of the energy curves of ground and excited states illustrating the libronic vertical downward transitions. Excited-state population probability is considered as exponential distribution or modified exponential distribution with suppressed probability of m = 0 state. b) Scheme of the lowest-energy libronic transitions coupled to the electronic $D_1 \rightarrow D_0$ transition with colours matching the peaks in the panels c-e. c-e) Comparison of experimental (black dots) and simulated (solid blue line) STM-EL fine spectra of $Q^+$ peak of $ZnPc^+$ and $MgPc^+$ (c,d) and $Q_{y1}^+$ and $Q_{y2}^+$ of $H_2Pc^+$ (e). The Franck-Condon factors are calculated including the modified exponential distribution and are colour-coded according to the vibration quantum number difference between the initial and final state, i.e. m - n (red - negative, blue - positive, grey - null). The energies $E_{00}$ of the 0-0 transition are set as the reference in each spectrum. Parameters of the simulated spectra are listed in Table 1.*

Qualitative comparison between the measured and simulated spectra using different combinations of the potential stiffnesses $k_1$, $k_0$ and the angular displacement $\Delta\phi_0$ (shown in Extended Data Fig. 3) yields an estimation of their physically relevant values for each studied case. Thus, for $\Delta\phi_0 \approx 0$ the simulation resembles the spectrum of $H_2Pc^+$, and for nonzero $\Delta\phi_0$ and a $k_1/k_0$ ratio above 1.1 approximates the progression observable in the spectra of $ZnPc^+$ and $MgPc^+$. Although the envelope and energetic distribution of the simulated peaks are in agreement with the experiments, the relative peak intensities for $ZnPc^+$, $MgPc^+$ cannot be well reproduced for any combination of parameters using the initial state population probability defined in Eq. (3). In particular, the overall intensity of the central peaks (near to the energy of 0-0 transition $E_{00}$) generated by the $m = n$ transitions is significantly lower than the one originating from $m - n = \pm 1$. This is an indication of an efficient depopulation channel for the lowest energy libron mode of the system in the excited state (*i.e.* the upper parabola in Fig. 3a). To reflect this in the theoretical model we modified the distribution $w'_m$ in Eq. (3) by reducing the probability of the zero-level libron excited state, using a factor $A < 1$ ($w'_0 = A \cdot w_0$) and fitted the experimental spectra (fitting parameters are summarized in Table 1).

| | $k_0$ (meV/(°)²) | $k_1$ (meV/(°)²) | $\Delta\phi_0$ (°) | $T_{eff}$ (K) | A | $\gamma$ (meV) | $E_{00}$ (eV) |
|---|---|---|---|---|---|---|---|
| $ZnPc^+$ 4 ML | 1.64 | 1.84 | 0.603 | 62 | 0.5 | 0.64 | 1.5188 |
| $MgPc^+$ 4 ML | 1.41 | 1.69 | 0.698 | 75 | 0.69 | 0.95 | 1.5149 |
| $H_2Pc^+$ 3 ML ($Q^+_{y1}$) | 1.57* | 1.63* | 0.001 | 50* | 1* | 0.91 | 1.3940 |

*Table 1: Parameters of the simulated spectra in Figure 3c-e. Parameter values denoted by an asterisk (\*) for $H_2Pc$ were fixed; $k_0$ and $k_1$ were taken from the TD-DFT simulations and A was set to unity. For fitting of the $H_2Pc$ spectrum minor component, three additional free parameters characterizing the emission peak $Q_{y2}^+$ of the second tautomer were used: relative intensity I = 0.29, FWHM of dressing function $\gamma_2$ = 0.31 meV and mutual energy separation of the tautomers $\Delta E_{00}$ = 1.17 meV.*

$H_2Pc$ represents a case in which the molecule does not undergo any change of the equilibrium angle upon the excitation due to the symmetrical adsorption geometry. Therefore the $m \neq n$ transitions are mostly forbidden, except for the ones with non-negligible Franck-Condon factors resulting from the change in the stiffness of the potential ($k_1/k_0 \neq 1$). Conversely, due to non-negligible $\Delta\phi_0$, fitting of the rich spectral features on $ZnPc^+$ and $MgPc^+$ allow precise determination of the $k_0$, $k_1$, $\Delta\phi_0$ parameters (Fig.3 c, d) which are, given the simplifications used in our model, in excellent agreement with the values estimated from the calculations (see Extended Data Table 2 for comparison). The effective temperature above 50 K resulting from the fitting indicates that the excess energy of inelastic tunnelling electrons of a few eV can excite high librational states and create a transient initial state population above the zero-level libron state[37].

## Nanocavity tuning of the transition between zero-energy librations

In order to shed light on the nature of the depopulation mechanism that leads to diminished probability $w_0'$ and consequently suppressed central feature in the spectra of the asymmetrically adsorbed molecules, we measure how it is affected by opening/closing the nanocavity. Since it is known[38] that the effective lifetime of the excitation in the phthalocyanines is reduced by the confinement of the optical density of states (Purcell factor), we are expecting a modulation of $w_0'$ due to a variable radiative quenching by the nanocavity. The resulting dependence of the MgPc$^+$/4 ML-NaCl spectra on the tip-sample distance is presented in Fig.4a, along with the parameters $A(\Delta z)$, the central peak position $E_{00}(\Delta z)$ and the overall line broadening $\gamma$ in Fig.4b,c that we determined by fitting of each spectrum individually. At the first glance, a significant increase of the central peak intensity with decreasing tip-sample distance $\Delta z$ is apparent; there is also an overall redshift of the entire spectra (decreasing $E_{00}$) and peak broadening (increasing $\gamma$). The latter two can be attributed to the known Lamb/Stark shift resulting from the coupling of the excited states to the nanocavity[33]. However, the striking variation of the central peak intensity, with factor $A$ changing from 0.40 to 0.93, is a clear signature of suppressing the deexcitation channel due to the faster radiative rate induced by compressing the nanocavity. At the moment, the exact nature of the deexcitation channel responsible for the decreasing intensity of the 0-0 lines in ZnPc$^+$ and MgPc$^+$ is not fully clear to us. Nonetheless, it becomes evident that the competition between the radiative and nonradiative decay rates from the zero-vibration excited states to the vibrational ground state of the trion plays a crucial role.

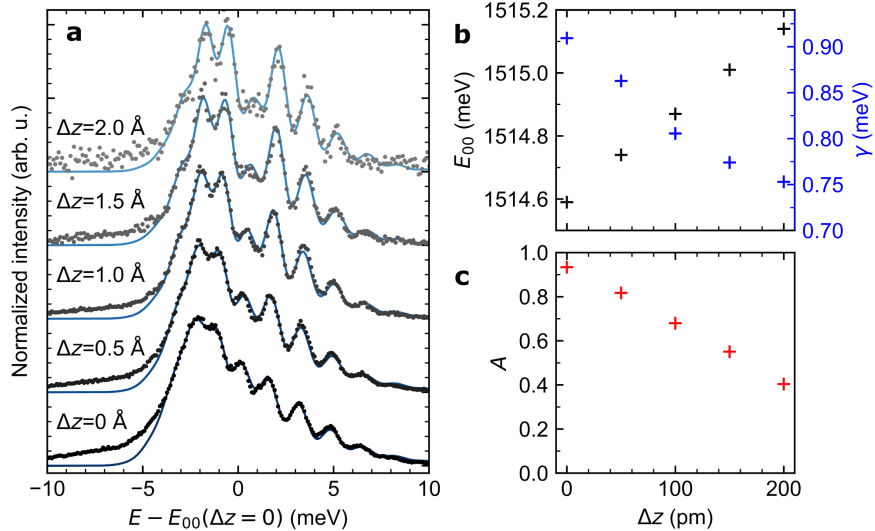

***Fig.4:*** a) Experimental (black dots) and simulated (solid blue lines) STM-EL fine spectra of Q$^+$ peak of MgPc$^+$ as a function of increasing tip-sample separation by $\Delta z$. V= -3 V, I ranges from 113 pA (bottom spectrum) to 19 pA (top spectrum). b, c) Central peak position $E_{00}$, linewidth $\gamma$ and central peak intensity A as a function of $\Delta z$ obtained from the fitting procedure of data in a).

*Average values of fitting parameters: $k_0$ = (1.69 ± 0.04) meV/(°)$^2$, $k_1$ = (1.42 ± 0.01) meV/(°)$^2$, $\Delta\phi_0$ = (0.69 ± 0.03)°, $T_{eff}$ = (82 ± 10) K.*

At this point, one may wonder if librational progressions could be also observed for the neutral excitons of the molecules. Based on the theoretical calculations on neutral molecules (see Extended Data Fig. 4 and Extended Data Table 2), revealing very similar values of $\Delta\phi_0$, $k_0$ and $k_1$ as the cations (Fig. 2), we believe the libration progressions are also present in the Q peak of neutral ZnPc and MgPc spectra, although indistinguishable due to a naturally broad energy character of the electroluminescence emission (see Extended Data Fig. 2). This might be overcome with a resonant STM-PL that already demonstrated its resolution capability on $Q_x$ of $H_2Pc$/4ML NaCl (0.5 meV linewidth)[20].

## Conclusions

To conclude, we found the link between the observation of libronic peak progression in the electroluminescence spectra and the chiral adsorption geometry of chromophores on NaCl. The molecules with chirally asymmetric adsorption configurations (Zn-, MgPc$^+$) change the orientation upon excitation, which according to the Franck-Condon principle allows the transitions between different libration states. This gives rise to the observed progressions in the spectra of chirally adsorbed chromophores. In contrast, in the reference system ($H_2Pc^+$) with a mirror-symmetrical adsorption geometry, the adsorption orientation remains unchanged upon excitation and, therefore, librational sidebands are not arising. From the analysis of the experimental spectra, it follows that the process of excitation gives rise to a non-equilibrium initial libration states population, corroborating one of the previously suggested mechanisms for the spectral broadening in STM-EL. Moreover, changes in the potential well stiffness of the libration, associated with the excitation, leads to an additional peak broadening. All these effects have to be considered for a correct interpretation of the STM-EL spectra of molecules in neutral and charged excited states. Finally, we have found experimental evidence of a depopulation pathway predominantly affecting the zero libration state of the trion. It can be effectively suppressed by overall increase of the radiative decay rate by closing the STM-EL nanocavity. We anticipate that the newly emerging methodology of STM-PL[20], enhanced with pump-probe capability could provide insight into the dynamics and physical origin of this deexcitation mechanism.

## Methods

**Sample preparation and STM measurements.** All measurements were performed in ultrahigh vacuum low temperature (at 7.5K) STM/AFM microscope with base pressure below $5\times10^{-10}$ mbar. The Ag(111) clean surface was prepared by standard cycles of sputtering and annealing. NaCl was evaporated from a source at 607°C on the surface kept at 120°C during 3-5 min to obtain 2-4 ML thick islands. Once the sample was inserted into the microscope head and cooled, the phthalocyanine molecules (from Sigma Aldrich) were deposited on it from a Ta crucible at 331°C ($H_2Pc$), 380°C (MgPc) and 415°C (ZnPc). We used a PtIr tip, sharpened by a focused-ion beam before inserting it into the scanner. Tips were cleaned and coated by Ag (or eventually Au) material by applying voltage pulses and controlled nanoscopic indentations into the clean substrate in order to achieve a suitable near-infrared plasmonic response.

**STML measurements.** In the optical spectroscopy measurements, we have used an optical setup described previously[27]. Overview and high-resolution photon spectra were obtained using gratings of 150 and 1200 grooves/mm and 100 µm wide slit, which provide spectral resolutions of 1.2 and 0.2 nm respectively corresponding to the best achievable energy resolution of 300, 400 and 600 µeV for $H_2Pc$ $Q^+_y$, $Q^+$ and Q of Mg/ZnPc peaks respectively.

**DFT and TD-DFT calculations.** The single-molecule calculations were run on $H_2Pc$, ZnPc and MgPc in both their neutral and cationic state. The ground state molecular structure was optimized in vacuo with density functional method (DFT) and ωB97X-D/6-31G* level of theory.[39] The emission properties are obtained by optimization of the first (and second for $H_2Pc$) excited states with TD-ωB97X-D/6-31G* for neutral and TDA-ωB97X-D/6-31G* for cations.[40] To calculate the total energy as a function of the molecular adsorption orientation, the electrostatic field of NaCl surface has been modelled as a slab of 3 layers of 6x6 point charges (+1 for Na+ and -1 for Cl-) with fixed position, at 3 Å of distance with the molecular plane. The optimized structure in vacuo of the neutral or cation molecule has been employed (B3LYP/6-31G*) and all coordinates are kept fixed except for the azimuthal angle $\phi$, varying between 0 and 45° (step of 1°) for ZnPc or MgPc, and between 0 and 90° (step of 5°) for $H_2Pc$. All calculations were performed with the Gaussian16 package[41].

**Spectroscopic fitting procedure.** Moment of inertia about the z-axis of the molecules was calculated from the ground state cation optimized structure (B3LYP/6-31G*). The resulting $J_z$ values are 113.7 $m_p$nm$^2$ for ZnPc$^+$, 114.3 $m_p$nm$^2$ for MgPc$^+$ and 113.2 $m_p$nm$^2$ for $H_2Pc^+$ where $m_p$ is the proton mass.

Spectrum of ZnPc$^+$ molecule (Fig. 3c) was fitted in the energy range ($E_{00}$-5 meV, $E_{00}$+10 meV) and spectra of MgPc$^+$ molecules (Fig. 3d, Fig 4a) were fitted in the energy range ($E_{00}$-4 meV, $E_{00}$+10 meV). The spectrum of $H_2Pc^+$ (Fig. 3e) was fitted in the range ($E_{00}$±4 meV). We used an iterative procedure minimizing the sum of least differences. Lorentzian dressing function corresponding to homogeneous broadening was used for the fitting of $H_2Pc^+$ spectrum. For $ZnPc^+$ and $MgPc^+$ spectra, we used the Gaussian as a dressing function. This can be rationalized by considering an inhomogeneous broadening caused by the NaCl relaxation upon molecular rotation between chiral adsorption configurations.

## Acknowledgements

SC, RCCF, MŠ and JD acknowledge the Czech grant agency funding no. 20-18741S and the Charles University Grant Agency project no. 910120. P.H. acknowledges support from project number L100101952 of the Czech Academy of Sciences. PM acknowledges grant EUR2021-122006 funded by MCIN/AEI/ 10.13039/501100011033 and, as appropriate, by "ERDF A way of making Europe", by the "European Union" or by the "European Union NextGenerationEU/PRTR". PM also acknowledges grant RYC2020-029800-I funded by MCIN/AEI/ 10.13039/501100011033 and, as appropriate, by "ESF Investing in your future" or by "European Union NextGenerationEU/PRTR".

## Author Contributions

J.D., R.C.C.F. and M.S. have conceived and performed the experiments and preprocessed the data for analysis. TD-DFT calculations and their analysis were performed by S.C. The model used for fitting was created by P.H., refined by J.D. and M.S. and the fitting was made by J.D. All authors have thoroughly discussed the data and contributed to the creation of the manuscript including figures and tables.

# Extended data figures and tables

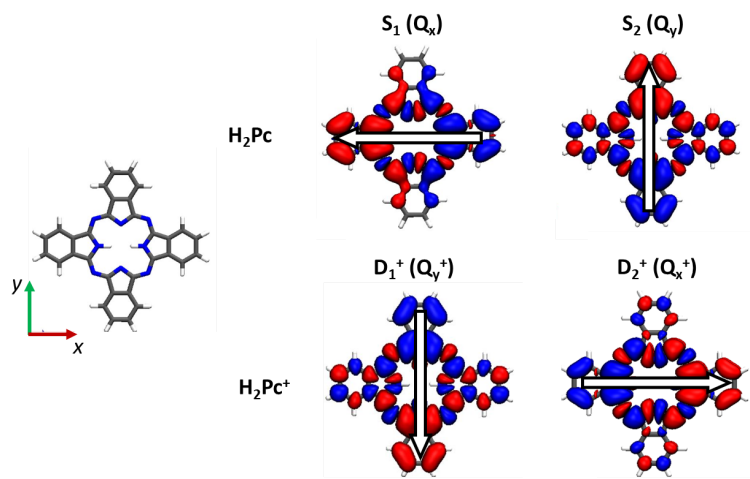

**Extended Data Fig. 1:** *Transition densities of the excited states of H$_2$Pc neutral and cation (see Extended Data Table 1). Calculations TD(TDA)-wB97XD/6-31G*. Isosurface 0.002 au.*

|  | n.state(sym) | Exc/eV (osc.) | band [a] | exp. |
|---|---|---|---|---|
| **H$_2$Pc neutral** | S$_1$ (B1u) | 1.89 (0.472) | Q$_x$ | 1.80 |
|  | S$_2$ (B2u) | 2.04 (0.542) | Q$_y$ | 1.93 |
| **H$_2$Pc+ cation** | D$_1$ (B2u) | 1.50 (0.241) | Q$_y^+$ | 1.39 |
|  | D$_2$ (B1u) | 1.71 (0.172) | Q$_x^+$ | - |
| **ZnPc neutral** | S$_1$ | 1.92 (0.510) | Q | 1.89 |
| **ZnPc+ cation** | D$_1$ | 1.63 (0.205) | Q$^+$ | 1.52 |
| **MgPc neutral** | S$_1$ | 1.88 (0.502) | Q | 1.89 |
| **MgPc+ cation** | D$_1$ | 1.59 (0.205) | Q$^+$ | 1.51 |

[a] For H$_2$Pc, *x* axis oriented along N-H .... H-N direction, see Extended Data Fig 1.

**Extended Data Table 1:** *Calculations TD- (TDA-)wB97XD/6-31G* of H$_2$Pc, ZnPc, MgPc neutral (cation): state number (with symmetry label), emission energy, oscillator strengths and band assignment. Comparison with experimental excitation energies taken from the spectra of Fig. 1.*

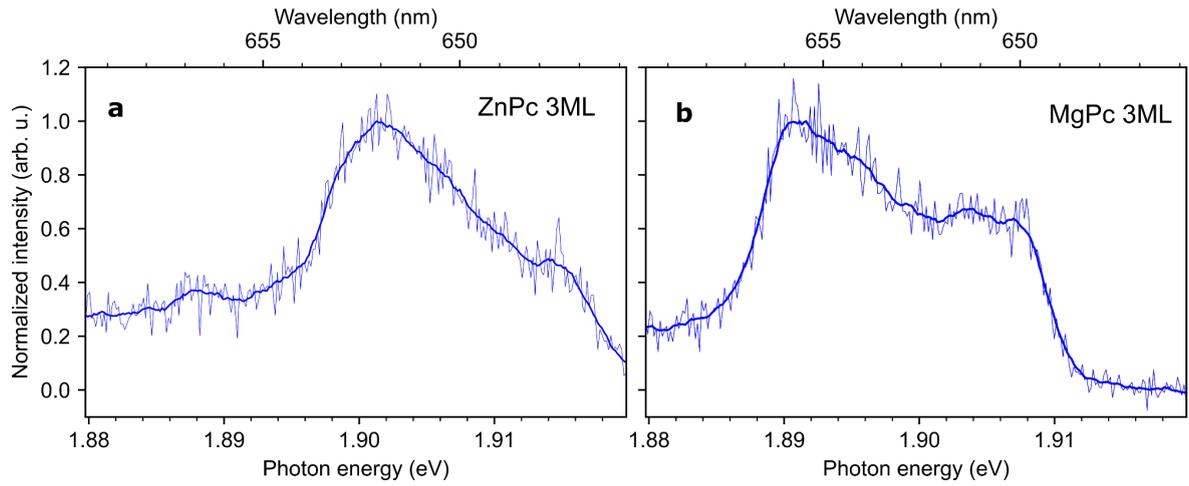

***Extended Data Fig. 2:*** *High-resolution STM-EL spectra measured at the lobe position corresponding to the neutral Q peak of ZnPc in a) and MgPc molecule in b) on 3 ML NaCl. Acquisition parameters: energy resolution 600 µeV, a) $U_S$ = -2.8 V , t = 60 s , I = 133 pA, b) $U_S$= -2.8 V, t = 60 s, I = 100 pA.*

|  | cation | | | neutral | | |
|---|---|---|---|---|---|---|
|  | $k_0$ (meV/(°)²) | $k_1$ (meV/(°)²) | $\Delta\phi_0$ (°) | $k_0$ (meV/(°)²) | $k_1$ (meV/(°)²) | $\Delta\phi_0$ (°) |
| ZnPc | 1.769 | 1.904 | 0.31 | 1.459 | 1.527 | 0.42 |
| MgPc | 1.747 | 1.880 | 0.28 | 1.461 | 1.535 | 0.41 |
| $H_2$Pc | 1.569 | 1.634 | 0.00 | 1.633 | 1.675 | 0.00 |

***Extended Data Table 2***: Parameters of parabolic fitting of computed total energies (see also Fig. 2 and Extended Data Fig. 4).

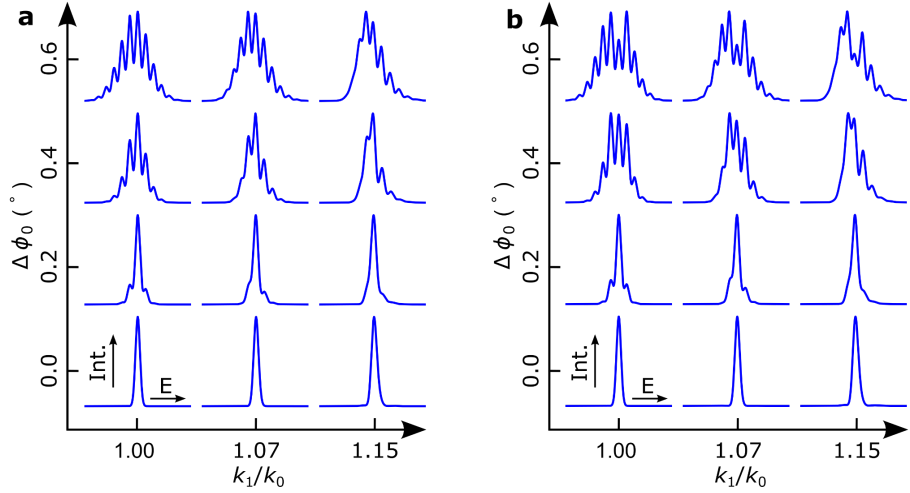

**Extended Data Fig. 3:** *Typology of simulated STM-EL spectra for varying values of excited/ground state equilibrium angle $\Delta\phi_0$, potential stiffness ratio $k_1/k_0$ and reduction factor A=1 in a) and A=0.5 in b) (using T = 70 K, $\gamma$ = 0.83 meV, $k_1$ = 1.83 meV/deg$^2$ and J = 113 $m_p$ nm$^2$). In the limiting case of $\Delta\phi_0$ = 0 and $k_1/k_0$ = 1, the spectrum consists of a single peak resulting from the sum of all m = n transitions. Increasing $\Delta\phi_0$ > 0 leads to significant overlap among different initial and final librational states (m - n ≠ 0) and consequent appearance of the red- and blue-shifted peak progressions. The ratio $k_1/k_0$ ≠ 1 affects the asymmetry of the spectral envelope around $E_{00}$ and the energy spacings among the peaks. For $k_1/k_0$ > 1, the interpeak energy difference in the red-shifted branch becomes smaller compared to the blue-shifted branch of the spectrum.*

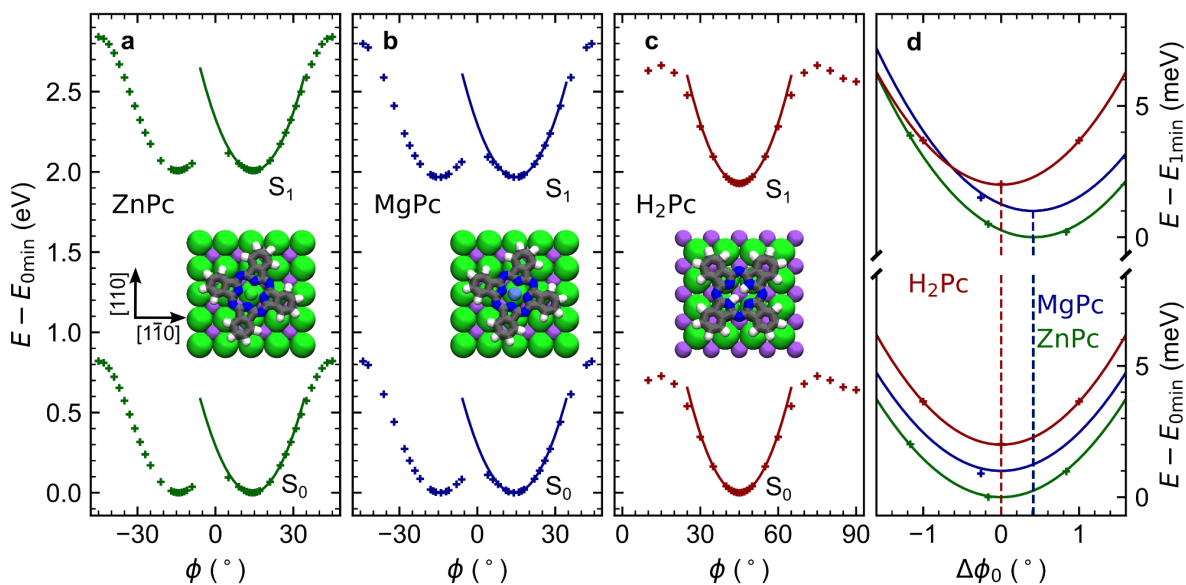

**Extended Data Fig. 4:** *Total energy as a function of rotation by angle φ for the ground and excited states of a) ZnPc, b) MgPc and c) H$_2$Pc: computed energy is plotted with points and the corresponding parabolic fits around the local minima with solid lines. The insets show the schematic models of the respective ground state neutral molecules in their equilibrium positions, i.e. rotated ~15° centred above Cl- in the case of ZnPc and MgPc, and rotated 45° above Na$^+$ in the case of H$_2$Pc. The angle φ is defined as between the molecule x-axes (crossing two opposing isoindole groups along N - N atom direction) and the [110] NaCl direction. d) The detailed comparison of the potential well minima of the three neutral chromophores as a function of the shift in the equilibrium angle positions Δφ$_0$ between the ground and excited states. MgPc and H$_2$Pc ground and excited state are vertically offset by 1 and 2 meV for clarity. Note the zero shift for H$_2$Pc, dictated by the symmetry of the system. Fitting parameters are summarized in Extended Data Table 2.*